\documentstyle[12pt]{article}
\begin{document}
\begin{center}
{\large \bf{$\pi $NN Coupling Constants
 from NN Elastic Data between 210 and 800 Mev}}
\vskip 1cm
D.V. Bugg
\vskip 0.6 cm
Queen Mary and Westfield College, London E1 4NS, UK
\vskip 0.6 cm
and
\vskip 0.6 cm
R. Machleidt
\vskip 0.6 cm
Department of Physics, University of Idaho, Moscow, Idaho 83843, USA
\end{center}
\skip 1.5 cm
High partial waves for $pp$ and $np$ elastic scattering are examined
 critically from 210 to 800 MeV.
Non-OPE contributions are compared with predictions from theory.
There are some discrepancies, but sufficient agreement that values of the
 $\pi NN$ coupling constants $g_0^2$ for $\pi ^0$ exchange and $g^2_{c}$ for
  charged $\pi $ exchange can be derived.
Results are $g^2_0 = 13.91 \pm 0.13 \pm 0.07$ and $g^2_c
 = 13.69 \pm 0.15 \pm 0.24$, where the first error is statistical and the
second is an estimate of the likely systematic error, arising mostly from
uncertainties in the normalisation of total cross sections and $d\sigma/
d\Omega$.

\newpage
\section {Introduction}
There has recently been controversy over the magnitude of the $\pi NN$
 coupling constant.
Prior to this controversy, the accepted value for many years was that
 determined by Bugg, Carter and Carter [1]: $f^2 = 0.0790(10)$,
where the error is given in parentheses. Defining
\begin {equation}
g^2 = f^2 (2M/\mu _{c})^2,
\end {equation}
where M is the mass of the proton and $\mu _{c}$ the mass of the charged
 pion, this value of $f^2$ corresponds to $g^2$ = $14.28(18)$.
This determination, based on the fixed $t$ dispersion relations for the
 $B^{(+)}$ amplitude in $\pi N$ elastic scattering, refers to
 exchange of charged pions at the nucleon pole: $\pi ^- p \rightarrow n$.
Since 1987, the Nijmegen group has challenged this value in a series
 of papers [2-4], analysing $NN$ elastic data up to 350 MeV with a potential
model.
 They find equal coupling constants for exchange of neutral and charged pions
 within experimental error: $f_0^2 = 0.0749(7)$ or $g^2_0 = 13.54(13),~f^2_c =
0.0741(5)$ or $g^2_c = 13.39(9)$, and a global average $f^2 = 0.0749(4)$ or
$g^2 = 13.54(7)$.

In 1990, Arndt et al. [5] obtained the value $f^2 = 0.0735(15)$,
 $g^2 = 13.28(27)$ from an analysis of new $\pi N$ data.
It transpires that they discarded much of the best but older data;
 a re-analysis,
 using all available data, and with careful attention to Coulomb effects, gives
 the result [6]: $f^2 = 0.0771(14),~g^2 = 13.94(25)$.
Meanwhile Machleidt and Sammarruca [7] have tested
 these results against precise information
 from the deuteron quadrupole moment
 and asymptotic D/S state ratio. They come down
 marginally in support of the higher values of $f^2$ and $g^2$.
At the recent Boulder conference, Arndt et al. [8] revised
 their determination upwards to $g^2 = 13.72(15)$, but stated that
 further analysis is still in progress, so this value should be considered
preliminary.

In this conflicting situation, it is relevant to examine  $NN$ elastic
 scattering data above the energy range considered by the Nijmegen group.
 These data have the virtue of being very precise and are a potential source
of information on $g^2_0$, the coupling constant for $\pi ^0$ exchange.
However, they do suffer from some difficulties.
A minor difficulty is inelasticity above 300 MeV, which introduces more free
 parameters into the phase shift analysis.
However, the serious issue is the separation of OPE from exchange of $2\pi$,
 $3\pi$, $\rho $, $\omega$, etc.
These will be described here collectively as ``heavy boson exchange'' (HBE).
In this paper, the conclusion is reached that $NN$ data from 140 to 800 MeV do
 indeed provide a useful determinations of $g^2$.

Pion exchange contributes through the $\delta $ amplitude [9] of the form [10]
\newline
 ${\bf (\sigma _1.q)(\sigma _2.q)}/(t - \mu ^2)$, where
${\bf {q}}$ is transverse momentum and $t = -q^2$.
It  may be isolated using the spin dependence of $NN$
 elastic scattering.
It turns out that this particular amplitude is well measured  by
 Wolfenstein parameters $D_{NN}$ for elastic scattering [11] and by $K_{NN}$
and $K_{LL}$ for $ np$ charge exchange.
The basic idea is illustrated in Fig. 1.
At $0^{\circ}$, amplitudes $\beta \propto {\bf {(\sigma _1.n)(\sigma _2.n)}}$
 and $\delta \propto \bf {(\sigma _1.q)(\sigma _2.q)}$
 are equal, since there is nothing to distinguish the normal
 $\bf {n}$ to the scattering plane and $\bf {q}$ in the plane of
 scattering.
The $\beta $ amplitude varies slowly with $q^2$ and can be extrapolated
 securely to $q = 0$.
The $\delta $ amplitude varies rapidly with $q$ and crosses zero at about
 $t = -\mu ^2$.
The difference between them determines the OPE amplitude.
Figs. 1(c) and (d) show that the parameters $D_{NN}$ and $K_{NN}$ have
 striking  dips and peaks respectively due to these zeros; these strong
features establish the magnitude of the OPE contribution.
A detail is that the Coulomb amplitude dominates pp scattering below
 $q \simeq 40$ MeV/c, so
for $pp$ the vital $q$ range is approximately 50 to 250 MeV/c.

A direct analysis of data on Wolfenstein parameters turns out to be difficult,
 because of interferences and because it is necessary
 to describe the slowly varying components in a way
 consistent with the HBE contributions, which are partly determined by other
 data. We therefore use the full weight of partial wave analysis and the
full data base.
The OPE contribution is then derived from the tensor components in high
 partial waves.
These are of two sorts: (a) mixing parameters $\overline \epsilon$,
 which mix triplet states having $J = L \pm 1$, and
 (b) tensor combinations of $H$ and $I$ waves [12]:
\begin {eqnarray}
H_T &\propto & -6^3H_4 + 11^3H_5 - 5^3H_6  \\
I_T &\propto & -7^3I_5 + 13^3I_6 -6^3I_7.
\end {eqnarray}
 In whatever way the analysis is done,
 the essential difficulty is to separate the
 exchange of heavier mesons from OPE. Actually $\sigma$ and $\omega $ exchange
 contribute only to central and spin-orbit combinations, which are othogonal
 to tensor components.
This leaves $\rho$ and $3\pi$ exchanges, which contribute a slowly varying
 component to tensor combinations.
Because of their large masses, they contribute mostly to low partial waves,
 notably $^3P_0$, $^3P_1$ and $^3P_2$ [13],
where they may be determined phenomenologically.
It  is well known that $\rho$ exchange cuts off the $\pi$ tensor amplitude
below $r \simeq 1$ fm.

In this paper, we will use the HBE contributions as determined by the
``Bonn peripheral model''. This is an extension of the model for higher
partial waves developed in section 5 of Ref. [14] above pion-production
threshold. It uses the approach described in Appendix B (Model II) of
Ref. [13]. The $\rho$ coupling is taken from the work of
H\"ohler and collaborators [15] and
the correlated $2\pi$ S-wave contribution from Durso et al.
Only the $\omega $ coupling is adjusted so as to fit $^3F_4$ and $^1G_4$.
 The model also includes $\Delta$(1232)-isobars
 in intermediate states which are
 excited via $\pi$ and $\rho$ exchange and provide inelasticity as well as
additional intermediate-range attraction.

The program of this work is to examine critically  each  of the high partial
 waves, so as to  form an opinion of what is (or is not) understood, before
 passing judgement on $g^2$.
The predictions by the peripheral Bonn model
are in excellent agreement with $pp$ data for partial
 waves with $J \geq 4$, with the  exception of $^3H_4$.
Corrections can be made for this small discrepancy, leaving what appears to be
a
 reliable determination of neutral $\pi $ exchange and the corresponding
 coupling constant $g^2_0$.

For $np$ charge exchange, OPE is three times larger than for $pp$
 elastic scattering, so one might hope to derive precise values of the charged
 coupling constant $g^2_c$.
The experimental data on Wolfenstein parameters are as good for $K_{NN}$
 in  charge exchange as for $D_{NN}$ in $pp$ elastic scattering.
Unfortunately, theory is not in such good shape, and there are clear
 discrepancies between HBE predictions and experiment for $I = 0$ $G$ and $H$
 waves. We find that it is still possible to make a
 reasonably accurate determination of $g^2_c$ from $\overline \epsilon _5$
and higher partial waves.

Section 2 analyses $pp$ scattering and arrives at a  determination of $g^2_0$.
 Section 3 analyses  $np$  data and $g^2_{c}$. Section 4
 comments on systematic errors and
other determinations of $g^2$. Section 5 presents conclusions.

\section {High partial waves for pp scattering}
There have been extensive and very accurate measurements of Wolfenstein
 parameters at TRIUMF from 210 to 515 MeV, at PSI by the Geneva group up
 to 580 MeV and at LAMPF from 485 to 800 MeV.
Where they overlap, these experiments
 agree well, so one can have confidence in the data.
In fact, what is needed to determine $g^2$ is a product of Wolfenstein
 parameters and $d\sigma /d\Omega$, and the bigger problems reside in the
 latter, where absolute normalisation presents experimental difficulties.
 Here the optical theorem is some help, by relating the imaginary part of the
 spin averaged amplitude  to
the total cross section via the optical theorem.
A  precise determination of $g^2$ depends on
 high absolute accuracy in all these data.
Results will be compared at eight energies from 210 to 800 MeV.
Because data come from a variety of experiments
using different techniques, one gets
 some idea of systematic errors.
At the Gatchina energy of 970 MeV, there are no accurate
measurements at the small angles required for present purposes.
Likewise, around 142 MeV, there are no measurements of Wolfenstein
 parameters below $30^{\circ}$. So these two energies are omitted.

Table 1 compares predictions for mixing parameters $\overline \epsilon _4$ and
 $\overline \epsilon _6$ with phase shift analysis. For $\overline
 \epsilon _4$, agreement is remarkably good right up to 800 MeV.
At the upper energies, HBE contributions are becoming uncomfortably large.
However, in
 view of the agreement for $\overline \epsilon _4$, one can have great
 confidence in  $\overline \epsilon _6$, where HBE contributions are much
 smaller.
Free fits to $\overline \epsilon _6$ at 580 and 800 MeV give satisfactory
 agreement with predictions.

The agreement for $\overline \epsilon _4$ is so good that one might wonder
 whether this parameter has already been used in optimising theoretical
predictions.
This is not so. We stress again that except for the $\omega$
(which does not create any tensor force and thus does not contribute
directly to $\overline \epsilon _J$) all parameters in the theoretical
model are taken from independent sources and have not been
fitted to data. So the agreement of
$\overline \epsilon _4$ with prediction up to 800 MeV is a real success
and not a circular argument.

Table 2 makes similar comparisions for $^1G_4$ and $H$ waves.
For $^1G_4$, agreement is excellent, but the HBE contribution is large
 compared with OPE.

For $^3H_4$, there is a definite discrepancy beginning at 325 MeV.
This raises the question of how far to trust HBE predictions for $^3K_6$.
In a classical approximation, angular momentum $L$ is related to impact
 parameter $r$ and momentum $k$ by $\sqrt {L(L + 1)} = kr$.
This suggests that agreement for $3H_4$ up to 210 MeV implies agreement for
 $^3K_6$ up to a lab energy T = 210 $\times (6 \times 7)/(4 \times 5) \simeq
 450$ MeV.
Calculations of HBE support this rough classical  notion for high partial
waves.
Using this prescription, the discrepancies for $^3H_4$ have been used to
 estimate small corrections for $^3K_6$ from 515 to 800 MeV, Table 3.
In practice, it turns out that this refinement has an effect on $g^2$ rather
 below statistical errors.

For $^3H_5$ and $^3H_6$, agreement is satisfactory up to 580 MeV; above this
 energy, small systematic discrepancies begin to appear.
It implies that $^3K_7$ and $^3K_8$ should be reliable up to 800 MeV
and this appears to be so  experimentally within two standard deviations.

Table 4 summarises the measure of agreement in the various partial waves.
A tick indicates agreement, a cross disagreement, L indicates that the HBE
 contribution is too large for comfort (a subjective judgement) and C
 indicates that a correction has been applied.

Table 5 gives values of $g^2_0$, depending on a variety of assumptions.
In the first column, $\overline \epsilon _4$ and $ ^3H_5$ are used,
 together with all parameters from $\overline \epsilon _6$ upwards.
 (At 210 MeV, $^3H_4$ is also used).
In Column 2, $\overline \epsilon _4$ is used but $H$ waves are left
 free.
In Column 3, only $\overline \epsilon _6$ and higher waves are used;
this is very conservative, almost certainly too conservative at 210 and 325
 MeV.

There is satisfactory consistency over most of Table 5.
In trying to derive a mean value for $g^2_0$,
it is necessary to steer a middle course between (a) using only the very high
partial waves, hence incurring a large error, or (b) risking that errors in HBE
affect $g^2_0$. We have chosen to  use HBE contributions
up to the energies where they become 20 - 25\% of OPE.
The reasoning is as follows.
We shall find errors on $g^2$ of about $\pm 2\%$.
It seems reasonable to believe HBE contributions to $10\%$ of their magnitudes
 in view of the excellent agreement for $\overline \epsilon _4$ (and later
$\overline \epsilon _3$ and $\overline \epsilon _5$).
This means that we use
$\overline \epsilon _4$, $^3H_5$, and $^1I_6$ up to 515 MeV.
We cut off $^3H_4$ above 210 MeV, because of the systematic discrepancies in
 Table 2.
It implies taking $g_0^2$ from column 1 of Table 5 up to 515 MeV and column 3
thereafter.
The result is a weighted mean
\begin {equation}
 g^2_0 = 13.91 \pm 0.13.
\end {equation}
The reader can easily make any other particular choice.

At 720 MeV, the data base is decidedly thin, and freeing $\overline \epsilon
_4$
 leads to considerable latitude in the phase shifts.
The value $g^2_0 = 10.81 \pm 0.83$ at this energy
in the third column of Table 5 is four standard
deviations from the mean and has therefore been discarded.

As stated in the Introduction, the OPE amplitude is being determined
essentially between $q$ = 50 and 250 MeV/c, i.e. at a mean value of
$t \simeq -\mu ^2$.
One has to worry about the effect of a form factor.
We write the OPE amplitude proportional to
\begin {equation}
\frac {g^2}{t - \mu ^2} \frac {\mu ^2 - \Lambda ^2}{t - \Lambda ^2} =
g^2 \left( \frac {1}{t - \mu ^2} - \frac {1}{t - \Lambda ^2} \right).
\end {equation}
The Bonn fit to lower partial-wave phase parameters requires $\Lambda$ = 1.3
 to 1.7 GeV/c$^2$ [13].
It is then straightforward to make a partial wave decomposition of the term
$g^2/(t - \Lambda ^2)$.
The result, using $\Lambda = 1.4$ GeV/c$^2$, is a perturbation to
 $\overline \epsilon _4$ of +0.04$^{\circ}$ at 800 MeV and less for lower
energies and higher partial waves.
This is completely negligible.
Physically it corresponds to the fact that the distant pole at $\Lambda$
 = 1.4 GeV/c$^2$ affects only low partial waves.
We remark, however, that the perturbation for $\overline \epsilon _3$ is
$-0.38^{\circ}$ at 800 MeV with $\Lambda$ = 1.4 GeV/c$^2$.
In the next section, we shall find excellent agreement between
$\overline \epsilon _3$ and experiment up to 650 MeV. This is an
independent check that $\Lambda$ cannot be substantially less than 1.4
Gev/c$^2$, since the correction varies roughly as $1/\Lambda ^2$.

We address the question of systematic errors on $g^2_0$ in Section 4.

\section {np Data and $g^2_c$}
 Table 6 compares predictions for $\overline \epsilon _3$ with experiment.
Agreement is satisfactory, except at 800 MeV, where prediction is dropping away
from experiment.
However, contributions from HBE are too large to allow  direct use of
$\overline
 \epsilon _3$ in determining $g^2_0$.
Nonetheless, one
 can have great confidence in using $\overline \epsilon _5$, which is also well
 determined experimentally.
Table 7 shows values of $\overline \epsilon _5$ from the phase shift
 analysis of Bugg and Bryan [16].

For $G$ and $H$ waves, the story is not so nice, Table 8.
Above 325 MeV, there are large discrepancies between experiment and Machleidt's
 predictions for $^3G_3$ and $^3G_4$ and a hint of disagreement for $^3G_5$.
The former discrepancies are certainly real.
 They are visible in TRIUMF $K_{SS}$,
 $K_{LS}$ and $A_{NN}$ data at 425 and 515 MeV and in independent
 LAMPF data for the same parameters at higher energies.
An experimental cross-check is that there is no
 apparent problem with $K_{NN}$ data, which were measured with the
 same technique and at the same time as $K_{SS}$ and $K_{LS}$.
Where they overlap near 500 MeV, TRIUMF and LAMPF data agree for $A_{NN}$ and
 $K_{NN}$.
We note that the fit to $^3G_4$ at higher energies could be improved by
a drastic reduction of $g^2_c$ (to 11.41 $\pm 0.19$ from 800 MeV data).
However, this would result in terrible predictions for $\overline \epsilon _3$
 and  $\overline \epsilon _5$.
Furthermore, $^3G_4$ at lower energies (T $\leq 325$ MeV) would
deteriorate substantially.
Thus there is no choice for $g^2_c$ that would consistently improve $^3G_4$
at all energies.
So the conclusion is that the disagreement must be taken seriously.

At 800 MeV, $^1H_5$ comes out significantly negative of OPE; $^1F_3$ shows the
same feature from 325 to 800 MeV [16].
Until the problem with the $G$ waves and $^1H_5$
is understood, one cannot have complete
 confidence in deriving $g^2_c$ from $ np$ data.

Table 9 summarises the measure of agreement between predictions
and $I = 0$ partial waves. From the impact parameter prescription [12],
 one can estimate perturbations to
 be applied to $I$ and $K$ waves for the discrepancies in Table 8.
The corrections, given in Table 3,
 are small and have the effect of lowering $g^2_c$ slightly above 515 MeV.
At 800 MeV, the experimental
determination of $^3I_6$ is compatible with the estimated
 correction.
At that energy, experiment is capable
 of determining partial waves up to $^3K_8$.

The result of this analysis is a determination of $g^2_c$ shown in Table 10
 using $\overline \epsilon _5$ and higher partial waves.
The weighted mean value of $g^2_c$ is
\begin {equation}
g^2_c = 13.67 \pm 0.29,
\end {equation}
where the error has been inflated from the statistical value $\pm 0.154$
 in order to cover fluctuations about the mean.
Corrections for the pion form factor are again completely negligible.

In view of the excellent agreement between $\overline \epsilon _3$ and
prediction, an alternative procedure is to use only $\overline \epsilon _5$,
which is extremely well determined by $K_{NN}$ data and also by $K_{LL}$
from 485 to 800 MeV. We mention in particulaar the data of Abegg et al. [17]
which lead to a very tight constraint on $\overline \epsilon _5$. Using this
 parameter alone leads to
\begin {equation}
g^2_c = 13.69 \pm 0.15.
\end {equation}
This value has the virtue of being independent of the small corrections
applied to $^1H_5$, $^3I_5$ and $^3I_6$.
It has a smaller error than equn. (6) because the fluctuations in
$\epsilon _5$ are smaller than those in higher partial waves.
It is our preferred value.
However, the error in equn. (7) is purely statistical and we shall show
in the next section that systematic errors are likely to be larger.

\section {Systematic errors}
The errors discussed so far are statistical. We now estimate systematic errors
arising from $d\sigma /d\Omega$ and Wolfenstein parameters, which we know to
be decisive in determining $g^2$.
For $pp$, it turns out that they are about $\pm 0.07$ for $g^2_0$,
i.e. slightly less than statistical errors.
For $g^2_c$ the situation is the reverse. Because of systematic uncertainties
 in normalisation, there is a systematic uncertainty which we tentatively
estimate as $\pm 0.24$.

For $pp$ scattering, the OPE amplitude is determined mostly by $D_{NN}d\sigma
/d\Omega$, where in conventional notation [9]
\begin {equation}
D_{NN} = \frac {|\alpha |^2 + |\beta |^2 - |\delta |^2 - |\epsilon |^2 +
2|\gamma |^2}{|\alpha |^2 + |\beta |^2 + |\delta |^2 + |\epsilon |^2 +
|\gamma |^2}.
\end {equation}
The measurement of $D_{NN}$ is made using a polarimeter which is calibrated
directly using the polarised beam. At the very small momentum transfers
relevant to OPE, any normalisation errors cancel, so this source of systematic
 errors can probably be neglected. At the most pessimistic, it should be
$\leq \pm 1\%$.

 The larger problem lies in $d\sigma /d\Omega$. Here we estimate a systematic
 error of $\pm 1\%$, for reasons we shall now detail.
 From 500 to 800 MeV, recent measurements of Simon et al. [18] have
statistical errors of $\pm 0.5\%$ and an absolute accuracy of about $\pm
0.5\%$.
 From 500 to 580 MeV, there are similarly precise data of Chatelain et al.
[19],
 and from 300-500 MeV precise 90$^{\circ}$ data of Ottewell et al. [20] with
normalisation errors of $\pm 1.8 \%$.
There are many measurements of the shape of $d\sigma /d\Omega$, but with
larger normalisation errors.
These can all be tied together with the $pp$ total cross section data of
Schwaller et al. [21] from 179 to 555 MeV, having systematic errors of
$\pm 0.8\%$ and statistics of $\pm 1\%$.
One finds in phase shift analysis that there are no particular conflicts
amongst all these data, so we conclude with an overall impression of a
$\pm 1\%$ error in $d\sigma /d\Omega$.

As regards OPE, the formula for $D_{NN}$
 depends essentially on $|\beta |^2 - |\delta |^2$,
so a $1\%$ systematic error translates into a $\pm 0.5 \%$ error in the
scale of the OPE amplitude, i.e. an uncertainty in $ g^2_0$ of $ \simeq \pm
0.07$.
This estimate has been checked by dropping in turn various sets of data from
the phase shift analysis and deliberately altering normalisations.

For $np$ charge exchange, the question of absolute normalisation has been
reviewed recently by McNaughton et al. [22].
For Wolfenstein parameters, the normalisation error is estimated at $\pm
1.8\%$.
However, the situation for $d\sigma /d\Omega$ is less satisfactory.
There are few measurements aiming at good absolute normalisation.
The difficulty lies in knowing the absolute flux of the neutron beam.
Keeler et al. [23] took great pains over this and claim an absolute
accuracy of $\pm 1.6\%$. However, even they admit to uncertainties of
$\pm 3\%$ for the relative normalisations of forward scattering
(where the neutron is detected) and charge exchange (where the proton is
detected).
Carlini et al. [24] present data at 800 MeV with good
absolute normalisation in the forward hemisphere.

The situation is confused by substantial discrepancies over $np$
total cross sections. These disagree amongst themselves and also disagree
with the differential cross sections we have just described.
Lisowski et al. [25] present data from 40 to 770 MeV with very high statistics;
they claim absolute normalisation of $\leq \pm 1\%$. Unfortunately their data
differ
by $6\%$ from most other data. These other data, although of lower
statistical accuracy, were mostly measured with monoenergetic neutron
beams and liquid hydrogen targets, whereas Lisowski et al. worked with
$CH_2 - C$ difference and a continuous neutron spectrum, a less attractive
technique as regards absolute normalisation.
Our phase shift solutions settle midway between the Lisowski et al. results and
the rest, but can be driven to fit either without too great a penalty in
$\chi ^2$.
Keeler et al. $d\sigma /d\Omega$ data definitely show a preference against
the Lisowski et al. data. If the latter are dropped, $g^2_c$ rises
systematically at all energies and averages to 13.90.
It therefore seems essential to allow $\pm 3\%$ normalisation
uncertainty for $d\sigma /d\Omega$.
Adding in quadrature the $\pm 1.8 \%$ uncertainty in $K_{NN}$, we arrive at
$\pm 3.5\%$ normalisation error in $|\beta |^2 - |\delta |^2$,
i.e. $\pm 1.75\%$ systematic uncertainty in the OPE amplitude.
This translates to a systematic error in $g^2_c$ of $\pm 0.24$.

We now comment briefly on other determinations of $g^2$.
The Nijmegen group has reported extensively on determination of $g^2$ from
 data up to 350 MeV.
There are two comments to be made on this work.
Firstly, it must contain similar uncertainties about HBE buried in
 the short-range potentials, but different in detail to those appearing here.
Secondly, it is interesting that they assert strongly that their values refer
 to the pion pole at $t = \mu ^2$.
Some years ago, forward dispersion relations were used [26] to find $g^2_0$
 from the $u$ channel pole below threshold.
This work was repeated with later data by Grein and Kroll [27].
It seems likely that the Nijmegen work is likewise finding $g^2_0$ from the
 $u$ channel contribution.
Their potential model calculation, via the Schr\"odinger equation, will
 include rescattering which appears in dispersion integrals.
The pion pole is only 10 MeV below threshold and makes its dominant
 contribution at low energies, through the energy dependence of $^1S_0$ and
$^3S_1$.
The Nijmegen group indeed finds that their greatest sensitivity is to data in
 the 10-30 MeV range.
They observe no particular sensitivity to Wolfenstein parameters.

If this assessment of the origin of their determination of
$g^2_0$ is correct, there is a corollary.
One needs to be cautious about Coulomb corrections, which are large at
 low energies and have a dramatic effect on the scattering length.
The Nijmegen group has been fastidious about the long range component of the
Coulomb potential. However, one needs to ask whether short-range
 Coulomb corrections to the scattering length and effective range are
 adequately understood (for example those
due to mass differences between $u$ and $d$
quarks).

The second route for  determining $g^2_c$ is from $\pi N$ fixed $t$
 dispersion relations.
This method should be  more reliable for $g^2_c$ than using
 $NN$ data.
Firstly, it evaluates the pole value directly. The nucleon pole lies between
 the $s$ and $u$ channel regions, and is obtained by interpolation
rather than from an extrapolation.
Secondly, the relevant $B^{(+)}$ amplitude is determined primarily by $P_{33}$,
indeed by the width of the $P_{33}$ resonance.
This partial wave is determined easily and accurately by total cross section
 measurements up to 300 MeV, aided by $d\sigma /d\Omega$ and $A_{0N}$ which
 determine small partial waves [28].
Total cross section measurements are a well tested technique, not only for
 $\pi N$ scattering but for many other reactions; worldwide agreement is
 characteristically at the $1-2\%$ level between different groups.
Thirdly, Coulomb corrections cancel to first order in the $B^{(+)}$ amplitude,
 which is a symmetric combination of $\pi ^+p $ and $\pi ^- p$.
Nonetheless, caution is essential because the $P_{33}$ amplitude has a
 significant splitting of mass and width between $\Delta ^{++}$ and
 $\Delta ^0$ states and Coulomb barrier corrections are sizeable.
The charge dependence is accurately measured experimentally and the Coulomb
 corrections have been studied in great depth theoretically [29].
So it appears that Coulomb corrections in $\pi N$ can be handled accurately.

\section {Concluding remarks}
NN data give consistent determinations of both $g^2_0$ and $g^2_c$ using
 data from 140 to 800 MeV.
The essential source of the information lies in precise measurments of
 Wolfenstein parameters $D_{NN},~K_{NN}$ and $K_{LL}$, together with $d\sigma /
d\Omega$.
Our preferred values are
\begin {eqnarray*}
g^2_0 &=& 13.91 \pm 0.13 \pm 0.07,  \\
g^2_c &=& 13.69 \pm 0.15 \pm 0.24.
\end {eqnarray*}
The latter value increases to 13.90 if the total cross section data of
 Lisowski et al. are dropped.
 The choice of determinations in Tables 5 and 10 is somewhat subjective, but
the
 interested reader can easily make his or her own choice.
Results are consistent with absence of charge dependence.

The result for $g^2_0$ is significantly larger than that of the Nijmegen group,
 $g^2_0 = 13.54 \pm 0.13$.
It is close to the latest value $g^2_c$ from ref. [6] and marginally above
Arndt's latest value $g^2_c = 13.72 \pm 0.15$.
Further precise measurements of Wolfenstein parameters in the 140-300 MeV
 range, below the inelastic threshold, should allow even further improvement
 in accuracy for $g^2_0$, and such measurements are in progress at IUCF [30].
At these energies, HBE corrections to $\overline \epsilon _4$, $G$ and $H$
 waves are surely small and accurately determined from a global fit to low
 partial waves and higher energy data.
Errors of $\pm 0.005$ on both Wolfenstein parameters and $d\sigma /d\Omega$
 may be achievable and would determine $g^2_0$
 with an accuracy of about $\pm 0.06$.
For np measurements, prospects of further improvements are not good, because
 of the great difficulty of measuring $d\sigma /d\Omega$ absolutely in charge
exchange.

{\bf Acknowledgements.}
It is a pleasure to thank Dr. R.
 Arndt for valuable discussions about $NN$ and $\pi N$ phase shift analysis.
 We also thank Dr. J. J. de Swart for providing numerical details of the
fits to data by the Nijmegen group.
The present work was supported in part by the U.S. National Science
Foundation Grant No.~PHY-9211607.

\pagebreak
{\bf Figure Caption}\\
Fig. 1. Real parts of $\beta$ and $\delta$ amplitudes for (a) pp elastic
 scattering (b) np charge exchange at 515 MeV, (c) and (d) corresponding
 values of $D_{NN}$ and $K_{NN}$.
\newpage

\begin {table}[htp]
\begin {center}
\begin {tabular} {ccccccc}
T(MeV) & OPE    & HBE   & OPE+HBE & Experiment & Discrepancy & Parameter
 \\\hline
210    & -1.128 & 0.021 & -1.107  & -1.29 $\pm$ 0.09 & -0.18 $\pm $ 0.09 &
 $\overline \epsilon _4$  \\
325    & -1.614 & 0.112 & -1.502 & -1.47 $\pm$ 0.07 & 0.03 $\pm$ 0.07 \\
425    & -1.939 & 0.262 & -1.677 & -1.66 $\pm$ 0.04 & 0.01 $\pm$ 0.04 \\
515    & -2.179 & 0.455 & -1.724 & -1.74 $\pm$ 0.07 & -0.02 $\pm$ 0.07  \\
580    & -2.329 & 0.625 & -1.704 & -1.67 $\pm$ 0.06 & 0.03 $\pm$ 0.06  \\
650    & -2.473 & 0.827 & -1.646 & -1.44 $\pm$ 0.06 & 0.21 $\pm$ 0.06  \\
720    & -2.601 & 1.037 & -1.564 & -1.66 $\pm$ 0.04 & -0.09 $\pm$ 0.04  \\
800    & -2.732 & 1.276 & -1.456 & -1.42 $\pm$ 0.04 & 0.04 $\pm$ 0.04  \\\hline
210    & -0.332 & 0.000 & -0.332 & & & $\overline \epsilon _6$ \\
325    & -0.554 & 0.003 & -0.551  \\
425    & -0.717 & 0.010 & -0.707  \\
515    & -0.843 & 0.020 & -0.823  \\
580    & -0.918 & 0.031 & -0.887 & -0.99 $\pm$ 0.05 & -0.10 $\pm$ 0.05  \\
650    & -1.003 & 0.046 & -0.957  \\
720    & -1.075 & 0.063 & -1.022  \\
800    & -1.149 & 0.087 & -1.062 & -0.98 $\pm$ 0.04 & 0.08 $\pm$ 0.04
 \\\hline
\end {tabular}
\caption { Comparing predictions for
 $\overline \epsilon _4$ and $\overline \epsilon _6$
with experiment. Units are degrees.}
\end {center}
\end {table}
\begin {table}[htp]
\begin {center}
\begin {tabular} {ccccccc}
T(MeV) & OPE    & HBE   & OPE+HBE & Experiment & Discrepancy & Parameter
 \\\hline
210    & 0.700 & 0.280 & 0.980  & 1.07 $\pm$ 0.07 & 0.09 $\pm $ 0.07 &
 $^1G_4$ \\
325    & 0.905 & 0.761 & 1.666 & 1.68 $\pm$ 0.05 & 0.01 $\pm$ 0.05 \\
425    & 1.011 & 1.345 & 2.356 & 2.31 $\pm$ 0.05 & -0.04 $\pm$ 0.05 \\
515    & 1.074 & 1.990 & 3.064 & 3.02 $\pm$ 0.04 & -0.04 $\pm$ 0.04  \\
580    & 1.106 & 2.495 & 3.601 & 3.68 $\pm$ 0.06 & 0.08 $\pm$ 0.06  \\
650    & 1.132 & 2.964 & 4.098 & 4.26 $\pm$ 0.06 & 0.16 $\pm$ 0.06  \\
720    & 1.150 & 3.246 & 4.396 & 4.15 $\pm$ 0.09 & -0.25 $\pm$ 0.09  \\
800    & 1.164 & 3.332 & 4.496 & 4.79 $\pm$ 0.07 & 0.29 $\pm$ 0.07  \\\hline
580    & 0.542 & 0.292 & 0.834 & 0.87 $\pm$ 0.04 & 0.03 $\pm$ 0.04  &
 $^1I_6$ \\
800    & 0.620 & 0.531 & 1.151 & 1.24 $\pm$ 0.05 & 0.09 $\pm$ 0.05  \\\hline
210    & 0.309 & 0.027 & 0.336 & 0.31 $\pm$ 0.05 & -0.03 $\pm$ 0.05  & $^3H_4$
\\
325    & 0.539 & 0.061 & 0.600 & 0.49 $\pm$ 0.05 & -0.11 $\pm$ 0.05 \\
425    & 0.723 & 0.073 & 0.796 & 0.50 $\pm$ 0.05 & -0.29 $\pm$ 0.05 \\
515    & 0.875 & 0.052 & 0.927 & 0.37 $\pm$ 0.05 & -0.56 $\pm$ 0.05  \\
580    & 0.977 & 0.006 & 0.983 & 0.51 $\pm$ 0.04 & -0.47 $\pm$ 0.04  \\
650    & 1.079 & -0.083 & 0.996 & 0.59 $\pm$ 0.06 & -0.41 $\pm$ 0.06  \\
720    & 1.176 & -0.231 & 0.945 & 0.48 $\pm$ 0.06 & -0.46 $\pm$ 0.06  \\
800    & 1.279 & -0.481 & 0.798 & 0.49 $\pm$ 0.04 & -0.31 $\pm$ 0.04  \\\hline
580    & 0.314 & 0.049 & 0.363 & 0.40 $\pm$ 0.04 & 0.03 $\pm$ 0.04  &
 $^3K_6$ \\
800    & 0.444 & 0.055 & 0.499 & 0.42 $\pm$ 0.04 & -0.08 $\pm$ 0.04
 \\\hline
325    & -1.286 & 0.183 & -1.103 & -1.19 $\pm$ 0.06 & -0.09 $\pm$ 0.06
  & $^3H_5$
\\
425    & -1.645 & 0.333 & -1.312 & -1.36 $\pm$ 0.05 & -0.05 $\pm$ 0.05 \\
515    & -1.926 & 0.501 & -1.425 & -1.56 $\pm$ 0.06 & -0.13 $\pm$ 0.06  \\
580    & -2.109 & 0.635 & -1.474 & -1.58 $\pm$ 0.07 & -0.11 $\pm$ 0.07  \\
650    & -2.289 &  0.772 & -1.517 & -1.66 $\pm$ 0.08 & -0.15 $\pm$ 0.08  \\
720    & -2.454 & 0.870 & -1.584 & -1.28 $\pm$ 0.07 & 0.30 $\pm$ 0.07  \\
800    & -2.626 & 0.917 & -1.709 & -1.48 $\pm$ 0.07 &  0.22 $\pm$ 0.07
\\\hline
580    & -0.847 & 0.103 & -0.744 & -0.75 $\pm$ 0.07 & -0.01 $\pm$ 0.07  &
 $^3K_7$ \\
800    & -1.124 & 0.194 & -0.930 & -0.96 $\pm$ 0.07 & -0.03 $\pm$ 0.07
 \\\hline
210    & 0.121 & 0.063 & 0.184 & 0.20 $\pm$ 0.04 & 0.02 $\pm$ 0.04 & $^3H_6$
 \\
325    & 0.230 & 0.185 & 0.415 & 0.42 $\pm$ 0.04 &  0.00 $\pm$ 0.04

\\
425    & 0.323 & 0.331 & 0.654 & 0.83 $\pm$ 0.04 & 0.18 $\pm$ 0.04 \\
515    & 0.402 & 0.475 & 0.877 & 0.81 $\pm$ 0.02 & -0.07 $\pm$ 0.02  \\
580    & 0.456 & 0.587 & 1.043 & 1.05 $\pm$ 0.03 & 0.01 $\pm$ 0.03  \\
650    & 0.512 &  0.707 & 1.219 & 1.37 $\pm$ 0.05 & 0.15 $\pm$ 0.05  \\
720    & 0.564 & 0.823 & 1.387 & 1.64 $\pm$ 0.05 & 0.25 $\pm$ 0.05  \\
800    & 0.621 & 0.943 & 1.564 & 1.79 $\pm$ 0.02 &  0.23 $\pm$ 0.02  \\\hline
580    & 0.167 & 0.105 & 0.272 & 0.40 $\pm$ 0.06 & 0.13 $\pm$ 0.06  &
 $^3K_8$ \\
800    & 0.246 & 0.207 & 0.453 & 0.50 $\pm$ 0.02 & 0.04 $\pm$ 0.02
 \\\hline
\end {tabular}
\caption { Comparison of $G$, $H$, $I$ and $K$ waves (degrees) with
experiment.}
\end {center}
\end {table}
\begin {table}[htp]
\begin {center}
\begin {tabular} {ccccc}
T(MeV) & $^3K_6$ & $^1H_5$ & $^3I_5$ & $^3I_6$  \\\hline
210 & 0.004 & 0.004 & -0.030 & 0.032 \\
325 & 0.013 &-0.021 & -0.083 & 0.088 \\
425 & 0.026 &-0.086 & -0.130 & 0.149 \\
515 & 0.018 &-0.182 & -0.130 & 0.150 \\
580 & 0.007 &-0.283 & -0.105 & 0.121 \\
650 &-0.014 &-0.421 & -0.071 & 0.083 \\
720 &-0.026 &-0.590 & -0.036 & 0.041 \\
800 &-0.056 &-0.821 &  0.000 & 0.000 \\\hline
\end {tabular}
\caption { Corrected HBE values from an impact parameter prescription.
 Units are degrees.}
\end {center}
\end {table}
\begin {table}[htp]
\begin {center}
\begin {tabular} {ccccccccccc}
T(MeV) & $\overline \epsilon _4$ & $^3H_4$ & $^1G_4$ & $^3H_5$ & $^3H_6$ &
$\overline \epsilon _6$ & $^3K_6$ & $^1I_6$ & $^3K_7$ & $^3K_8$ \\\hline
210 & $\surd$ & $\surd$  & L & $\surd$ & L & $\surd$ & $\surd$ & $\surd$ &
$ \surd$ & $\surd$ \\
325 & $\surd$ & $\surd$  & L & $\surd$ & L & $\surd$ & $\surd$ & $\surd$ &
$\surd$ & $\surd$ \\
425 & $\surd$ & $\times$ & L & $\surd$ & L & $\surd$ & $\surd$ & $\surd$ &
 $\surd$ & $\surd$ \\
515 & $\surd$ & $\times$ & L & $\surd$ & L & $\surd$ &   C   & $\surd$ &
$\surd$ & $\surd$ \\
580 &   L   & $\times$ & L &   L   & L & $\surd$ &   C   &   L   & $\surd$ &
 $\surd$ \\
650 &   L   & $\times$ & L &   L   & L & $\surd$ &   C   &   L   & $\surd$ &
$\surd$ \\
720 &   L   & $\times$ & L &   L   & L & $\surd$ &   C   &   L   & $\surd$ &
$ \surd$ \\
800 &   L   & $\times$ & L &   L   & L & $\surd$ &   C   &   L   & $\surd$ &
 $\surd$ \\
\hline
\end {tabular}
\caption { Summary of the measure of agreement between HBE and experiment for
$pp$;
$ \surd$ indicates agreement, $\times$ disagreement,
 L indicates that HBE is uncomfortably large
 and C indicates that an empirical correction has been applied using data on
$^3H_4$.}
\end {center}
\end {table}
\begin {table}[htp]
\begin {center}
\begin {tabular} {cccc}
T(MeV) & Using $\overline \epsilon _4$ and $^3H_5$ & Using
 $\overline \epsilon _4$ &
$\overline \epsilon _6$ upwards  \\\hline
210 & 15.01 $\pm$ 0.46 & 15.79 $\pm$ 0.48 & 15.13 $\pm$ 0.75  \\
325 & 14.23 $\pm$ 0.33 & 13.99 $\pm$ 0.35 & 14.70 $\pm$ 0.57  \\
425 & 13.95 $\pm$ 0.26 & 13.41 $\pm$ 0.34 & 13.44 $\pm$ 0.38  \\
515 & 14.02 $\pm$ 0.21 & 14.61 $\pm$ 0.25 & 14.58 $\pm$ 0.34  \\
580 & 13.75 $\pm$ 0.35 & 13.74 $\pm$ 0.60 & 14.45 $\pm$ 0.68  \\
650 & 12.86 $\pm$ 0.44 & 12.17 $\pm$ 0.48 & 12.88 $\pm$ 0.52  \\
720 & 14.01 $\pm$ 0.45 & 14.10 $\pm$ 0.45 & 10.81 $\pm$ 0.83  \\
800 & 13.69 $\pm$ 0.16 & 13.44 $\pm$ 0.16 & 13.61 $\pm$ 0.22  \\\hline
\end {tabular}
\caption { Fitted values of $g^2_0$ from three assumptions for HBE.}
\end {center}
\end {table}
\begin {table}[htp]
\begin {center}
\begin {tabular} {cccc}
T(MeV) & OPE+HBE & Experiment & Discrepancy  \\\hline
142 & 4.829 & 4.63 $\pm$  0.13 & -0.20 $\pm$  0.13  \\
210 & 6.211 & 6.00 $\pm$  0.07 & -0.21 $\pm$  0.07  \\
325 & 7.445 & 7.36 $\pm$  0.07 & -0.08 $\pm$  0.07  \\
425 & 7.882 & 7.98 $\pm$  0.11 &  0.10 $\pm$  0.11  \\
515 & 8.004 & 7.99 $\pm$  0.13 & -0.01 $\pm$  0.13  \\
650 & 7.913 & 8.05 $\pm$  0.13 &  0.14 $\pm$  0.13  \\
800 & 7.635 & 8.17 $\pm$  0.12 &  0.52 $\pm$  0.12  \\\hline
\end {tabular}
\caption { Comparison of fitted values of $\overline \epsilon _3$ with HBE.
 Units are degrees.}
\end {center}
\end {table}
\begin {table}[htp]
\begin {center}
\begin {tabular} {cccccc}
T(MeV) & OPE & HBE & OPE+HBE & Experiment & Discrepancy  \\\hline
142 & 1.205 & -0.011 & 1.194 &   \\
210 & 1.905 & -0.026 & 1.879 & 1.83 $\pm$  0.07 & -0.05 $\pm$  0.07  \\
325 & 2.881 & -0.067 & 2.814 & 2.71 $\pm$  0.05 & -0.11 $\pm$  0.05  \\
425 & 3.555 & -0.119 & 3.436 & 3.34 $\pm$  0.08 &- 0.10 $\pm$  0.08  \\
515 & 4.060 & -0.183 & 3.877 & 3.79 $\pm$  0.09 & -0.09 $\pm$  0.09  \\
650 & 4.681 & -0.313 & 4.368 & 4.37 $\pm$  0.10 &  0.00 $\pm$  0.10  \\
800 & 5.234 & -0.509 & 4.725 & 4.86 $\pm$  0.11 &  0.13 $\pm$  0.11  \\\hline
\end {tabular}
\caption { Comparison of fitted values of $\overline \epsilon _5$ with HBE.
 Units are degrees.}
\end {center}
\end {table}
\begin {table}[htp]
\begin {center}
\begin {tabular} {ccccccc}
T(MeV) & OPE & HBE & OPE+HBE & Experiment & Discrepancy & Parameter
\\\hline
142 & & & -1.676 & & & $^3G_3$ \\
210 & & & -2.860 & -3.20 $\pm$ 0.16 & -0.34 $\pm$ 0.16 &  \\
325 & & & -4.647 & -3.95 $\pm$ 0.13 &  0.70 $\pm$ 0.13 & \\
450 & & & -5.835 & -5.26 $\pm$ 0.19 &  0.58 $\pm$ 0.19 & \\
515 & & & -6.593 & -5.84 $\pm$ 0.17 &  0.75 $\pm$ 0.17 & \\
650 & & & -7.222 & -6.96 $\pm$ 0.20 &  0.26 $\pm$ 0.20 & \\
800 & & & -7.346 & -6.38 $\pm$ 0.16 &  0.97 $\pm$ 0.16 & \\\hline
800 & -2.278 & -0.091 & -2.369 & -2.20 $\pm$ 0.10 & 0.17 $\pm$ 0.10 & $^3I_5$
 \\\hline
142 & 3.263 & 0.209 & 3.472 & & & $^3G_4$ \\
210 & 4.931 & 0.427 & 5.358 & 5.84 $\pm$ 0.14 & 0.48 $\pm$ 0.14 & \\
325 & 7.241 & 0.776 & 8.017 & 7.84 $\pm$ 0.20 &-0.18 $\pm$ 0.20 & \\
425 & 8.856 & 0.934 & 9.790 & 8.71 $\pm$ 0.15 &-1.08 $\pm$ 0.15 & \\
515 &10.081 & 0.900 &10.981 & 9.67 $\pm$ 0.16 &-1.31 $\pm$ 0.16 & \\
650 &11.618 & 0.496 &12.114 &10.80 $\pm$ 0.20 &-1.31 $\pm$ 0.20 & \\
800 &13.014 & -0.433&12.581 &10.20 $\pm$ 0.18 &-2.38 $\pm$ 0.18 & \\\hline
800 & 5.199 & 0.300 & 5.499 & 5.25 $\pm$ 0.16 & -0.25 $\pm$ 0.16 & $^3I_6$
 \\\hline
142 &-0.467 & 0.195 & -0.272 & & & $^3G_5$ \\
210 &-0.825 & 0.454 & -0.371 & & & \\
325 &-1.410 & 1.031 & -0.379 & -0.52 $\pm$ 0.20 &-0.14 $\pm$ 0.20 & \\
425 &-1.872 & 1.609 & -0.263 & -0.70 $\pm$ 0.14 &-0.44 $\pm$ 0.14 & \\
515 &-2.250 & 2.158 & -0.092 & -0.36 $\pm$ 0.17 &-0.27 $\pm$ 0.17 & \\
650 &-2.756 & 2.999 &  0.243 &  0.11 $\pm$ 0.12 &-0.13 $\pm$ 0.12 & \\
800 &-3.245 & 3.901 &  0.656 &  0.18 $\pm$ 0.19 &-0.48 $\pm$ 0.19 & \\\hline
800 & -1.219 & 0.676 & -0.543 & -0.69 $\pm$ 0.07 & -0.15 $\pm$ 0.07 &
$^3I_7$  \\\hline
800 & -2.495 & -0.095 & -2.590 & -3.39 $\pm$ 0.09 & -0.80 $\pm$ 0.09 &
$^1H_5$ \\\hline
\end {tabular}
\caption { Comparison of fitted values of $I = 0$ $G$ and $H$ waves with HBE.
Units are degrees.}
\end {center}
\end {table}
\begin {table}[htp]
\begin {center}
\begin {tabular} {ccccccccccc}
T(MeV) & $\overline \epsilon _3$ & $^3G_3$ & $^3G4$ & $^3G_5$ &
$\overline \epsilon _5$ & $^3I_5$ & $^1H_5$ & $^3I_6$ & $^3I_7$ \\\hline
210 & $\surd$ & $\surd$  & ? & $\surd$ & $\surd$ & $\surd$ & $\surd$
 &$\surd$ & $\surd$ \\
325 & $\surd$ & $\times$ & $\surd$ & $\surd$ & $\surd$ & $\surd$ &   C   &
$\surd$ & $\surd$ \\
425 & $\surd$ & $\times$ & $\times$& ? & $\surd$ & $\surd$ &   C   &
$\surd$ & $\surd$ \\
515 & $\surd$ & $\times$ & $\times$& $\surd$ & $\surd$ &   C   &   C   &   C
  & $\surd$ \\
650 &$\surd$ & ? & $\times$& $\surd$& $\surd$ &   C   &   C   &   C
  & $\surd$ \\
800 &$\times$ &$\times$ & $\times$& $\times$& $\surd$ &   C   &   C   &   C
  & $\surd$ \\\hline
\end {tabular}
\caption { Summary of the measure of agreement between HBE and experiment
 for $I = 0$ phase shifts;
 $\surd$ indicates agreement, $\times$ disagreement, L indicates that HBE is
uncomfortably large,and C indicates that a correction has been applied to
HBE predictions using experimental data from G waves.}
\end {center}
\end {table}
\begin {table}
\begin {center}
\begin {tabular} {cc}
T(MeV) & $g^2_c$  \\\hline
142 & 13.13 $\pm$ 1.70 \\
210 & 11.84 $\pm$ 0.78  \\
325 & 13.99 $\pm$ 0.45  \\
425 & 12.68 $\pm$ 0.42 \\
515 & 14.46 $\pm$ 0.43  \\
650 & 14.23 $\pm$ 0.34  \\
800 & 13.47 $\pm$ 0.31  \\\hline
\end {tabular}
\caption { Fitted values of $g^2_c$.}
\end {center}
\end {table}
\end {document}